\documentclass[%
reprint, superscriptaddress,
showpacs,preprintnumbers,
dvipsnames,svgnames,table,
 amsmath,amssymb,
 aps,nofootinbib]{revtex4-1}
\usepackage{amsmath}
\usepackage{multirow}
\usepackage{tikz}
\usepackage{graphicx}
\usetikzlibrary{positioning}
\usepackage{dcolumn}
\usepackage{bm}

\newcommand{\ket}[1]{|{#1}\rangle}
\newcommand{\bra}[1]{\langle{#1}|}

\newcommand{\be}{\begin{equation}}
\newcommand{\ee}{\end{equation}}
\newcommand{\ba}{\begin{eqnarray}}
\newcommand{\ea}{\end{eqnarray}}
\newcommand{\ban}{\begin{eqnarray*}}
\newcommand{\ean}{\end{eqnarray*}}

\usepackage[colorlinks]{hyperref}
\usepackage[caption=false]{subfig}
\usepackage{float}
\usepackage{epstopdf}
\usepackage{enumitem}
\usepackage{subfig}

\begin{document}

\title{Power of an optical Maxwell's demon in the presence of photon-number correlations}%

\author{Angeline Shu}
 \affiliation{Department of Physics, National University of Singapore, 2 Science Drive 3, Singapore 117542.}
 \affiliation{Centre for Quantum Technologies, National University of Singapore, 3 Science Drive 2, Singapore 117543.}
\author{Jibo Dai}
\email{cqtdj@nus.edu.sg}
 \affiliation{Centre for Quantum Technologies, National University of Singapore, 3 Science Drive 2, Singapore 117543.}
\author{Valerio Scarani}
 \affiliation{Department of Physics, National University of Singapore, 2 Science Drive 3, Singapore 117542.}
 \affiliation{Centre for Quantum Technologies, National University of Singapore, 3 Science Drive 2, Singapore 117543.}

\date{\today}

\begin{abstract}
We study how correlations affect the performance of the simulator of a Maxwell's demon demonstrated in a recent optical experiment [Vidrighin et al., Phys. Rev. Lett. \textbf{116}, 050401 (2016)]. The power of the demon is found to be enhanced or hindered, depending on the nature of the correlation, in close analogy to the situation faced by a thermal demon.

\end{abstract}

\maketitle


\section{Introduction}

The Maxwell's Demon, first introduced in a thought experiment by James Clerk Maxwell \cite{Maxwell:71}, is a being with the ability to extract work from a system in contact with a single thermal bath, in apparent violation of the second law of thermodynamics. Since everyone believes that the second law is not to be violated, a long series of exorcisms of the demon have been proposed \cite{Leff+Rex:03,Maruyama+2:09}. Following Landauer \cite{Landauer:61} and Bennett \cite{Bennett:82}, today there is a broad consensus that information must come into the balance, and specifically that information erasure comes with an entropy cost; though some discordant voices remain (see e.g.~\cite{Alicki:14} and references therein). In the last decade or so, it was noticed that the most powerful demon should be able to manipulate information at the quantum level \cite{Zurek:03, Kim+3:11, Chapman+Miyake:15}. Studies of information balances have provided intriguing insights \cite{Chapman+Miyake:15, Delrio+4:11,Braga+3:14}, but the connection with usual thermodynamics requires a quantitative definition of work, which is a subject of controversy in the quantum regime \cite{Dahlsten+3:11,  Hovhannisyan+3:13, Skrzypczyk+2:14, Roncaglia+2:14, Gallego+2:16,  Talkner+Hanggi:16, Woods+2:16}.

One way to sort out theoretical discussions is to resort to experiments \cite{Price+4:08, Toyabe+4:10, Berut+5:11, Koski+3:14, Vidrighin+5:16}. While such simulations of the Maxwell's demon cannot be used to draw conclusions about ultimate limits, they do provide a concrete setting in which to study the power of the demon. In the recent optical simulation by Vidrighin and coworkers \cite{Vidrighin+5:16}, a two-mode optical field impinges on two photodiodes and the electric charges thence emitted are used to charge a capacitor. In the limit of ideal linear photodiodes, every photon creates an electron: therefore the voltage is directly proportional to the difference of photon number between the two modes. Initially, the two modes carry independent fields with identical photon-number distribution (chosen as thermal). In the absence of the demon, the two photodiodes produce on average the same photocurrent and the capacitor is not charged. The demon is mimicked by a weak monitoring of the two fields, realised approximately by photon subtraction and a single photon detection. Some detection patterns imply a bias in the number of photon at a given time, which can be used to charge the capacitor.

\begin{figure}[b!]		
\centering
\subfloat{}{\includegraphics{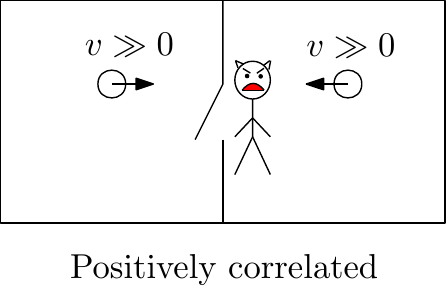}}\vspace{2ex}
\subfloat{}{\includegraphics{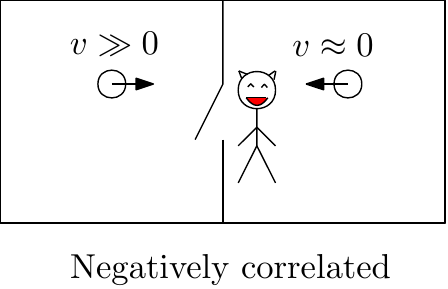}}
\caption{For the original ``thermal'' demon, one doesn't expect correlations between the fluctuations of the two halves. But were one to play that game for the sake of it, the effect of the correlations on the demon's power is intuitive. If at the trapdoor, each time a fast molecule comes from the left there is a fast molecule coming from the right, the demon's action will be hindered. If each time a fast molecule comes from the left there is a slow molecule coming from the right, the power of the demon is enhanced.}
\label{fig:intuitive}
\end{figure}

In this paper, we focus on this setup to investigate how the power of the demon varies by changing the photon-number statistics of the two fields. We mostly consider cases in which the partial state of each mode is still thermal, but the photon numbers may be correlated between the modes (not with the demon). We shall notably see that the behavior of the simulation is analog to the one expected for the thermal demon (see Fig.~\ref{fig:intuitive}).

\section{Framework for the study}

\subsection{The setup}

\begin{figure}[b!]		\centering\includegraphics[width=0.4\textwidth]{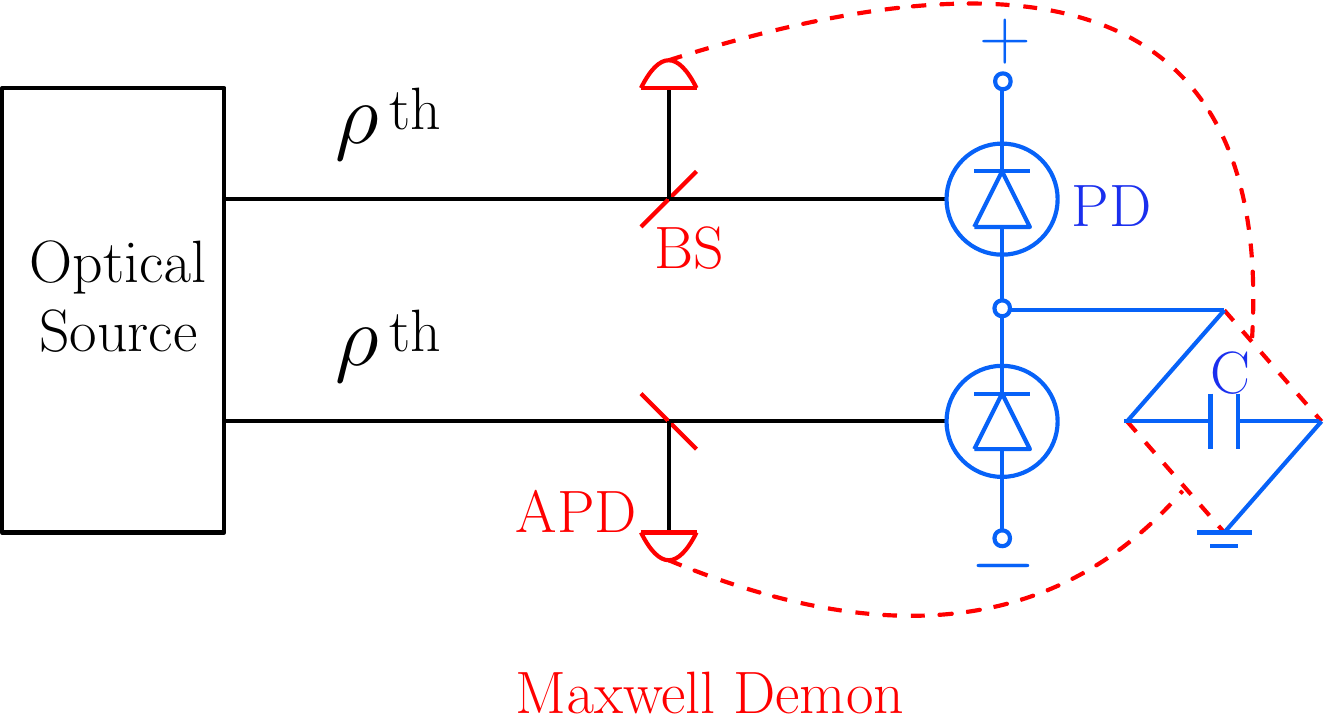}
\caption{(Color online) The setup under study, explained in detail in the text. It consists of the optical source and modes $\rho$, the electronic detection circuit with the linear photo-diode (PD) and capacitor (C), as well as the devices of the demon with the beam splitter (BS) and the avalanche photo-diode (APD). In the color version, the demon is in red, the optical elements in black and the electronic circuitry in blue.}
\label{figall}
\end{figure}

The measurement setup is sketched in Fig.~\ref{figall}. Two optical modes are sent on photodiodes operated in proportional mode, and the difference in photocurrents is used to charge a capacitor. If the goal were to optimise the charging of the capacitor, one would simply leave one of the modes in the vacuum state, thus achieving maximal bias. Mimicking the Maxwell's demon thought experiment rather requires the two modes to carry the same average number of photons $\bar{n}_{\mathrm{A}}=\bar{n}_{\mathrm{B}}$: thus, in the absence of the demon, there will be no net charging.

An even more accurate simulation of the demon demands that the states are uncorrelated \textit{thermal states} at the same temperature, as coming from two sources in contact with the \textit{same} bath. Besides, like in the experiment we keep only a mono-frequency component of the blackbody radiation\footnote{In the experiment, the field was mono-frequency because of the way the state was prepared: not by coupling to a bath, but by a customary randomisation of an initially coherent laser beam. For the present theory, it is clear that the demon simulator under study does not distinguish between frequencies, so keeping the whole blackbody spectrum would just add unnecessary complications to the toy model. Needless to say, a demon that could sort frequencies would be more powerful.}. A monomode optical field in equilibrium with a thermal bath at temperature $T$ is in the thermal state 
\begin{equation}
\rho^{\mathrm{th}}(\beta)=\left(1-\mathrm{e}^{-\beta\hbar\omega}\right)\sum_{n=0}^\infty\mathrm{e}^{-n\beta\hbar\omega}|n\rangle\langle n|,
\end{equation} with $\beta=1/(k_{\mathrm{B}}T)$ as usual. As well known, the average photon number in such a state is
\begin{equation}
\bar{n}=\frac{1}{\mathrm{exp}(\beta\hbar\omega)-1}=\frac{\lambda}{1-\lambda},
\end{equation}
with $\lambda=\mathrm{exp}(-\beta\hbar\omega)$.
The state prepared in the experiment was therefore
\begin{equation}\label{eqpassive}
\rho^{\mathrm{th}}_{\mathrm{AB}}=\rho^{\mathrm{th}}_{\mathrm{A}}\otimes\rho^{\mathrm{th}}_{\mathrm{B}}\,.
\end{equation}

The demon is implemented as follows: a beam-splitter is placed in front of each photodiode, and the reflected beams are monitored with photon counters. An intuitive picture of the demon's working can be gained in the regime, in which the reflectivity of the beam splitter is low (later in the paper this reflectivity will be a free parameter in an optimisation). In this regime, the demon realises a \textit{photon subtraction} in each beam \cite{Parigi+3:07, Zavatta+3:08}. If the initial state is $\rho$, upon recording a click by the photon counter in the reflected beam, the conditional state of the transmitted beam becomes the photon subtracted state, $\rho_\mathrm{sub}=a\rho a^\dagger/\mathrm{tr}\{a\rho a^\dagger\}$. The average number of photons in the photon-subtracted state $\bar{n}_\mathrm{sub}$ is related to the initial one $\bar{n}$ by $\bar{n}_\mathrm{sub}=\bar{n}-1+\sigma^2_n/\bar{n}$, with $\sigma^2_n$ being the variance of the initial photon number statistics \cite{Ueda+2:90}. For a thermal state, it holds $\bar{n}_\mathrm{sub}=2\bar{n}$, that is, the thermal state being super-Poissonian, the fact that a photon has been subtracted \textit{increases} the expected number of photons. In other words, for that state, if one of the counters clicks and the other does not, there are on average more photons in the mode whose detector has clicked. One can then choose the polarity of the capacitor accordingly and achieve a net charging.

\subsection{Tools for the calculation}

Since both the demon's operation and the final measurement are not sensitive to coherence in the number basis, without loss of generality we can study photon statistics coming from states that are diagonal in that basis:
\ba
\rho_{\mathrm{AB}}&=&\sum_{n_{\mathrm{A}},n_{\mathrm{B}}}p(n_{\mathrm{A}},n_{\mathrm{B}})\,\ket{n_{\mathrm{A}}}\bra{n_{\mathrm{A}}}\otimes\ket{n_{\mathrm{B}}}\bra{n_{\mathrm{B}}}\,.
\ea
On each mode $\mathrm{X}=\mathrm{A},\mathrm{B}$, the demon's operation consists of inserting a beam-splitter
\ba\label{beamsplitter}
\ket{n_\mathrm{X}}\bra{n_\mathrm{X}}&\longrightarrow&\sum_{k=0}^n{n \choose k}\,(1-R{_\mathrm{X}})^{n-k}R_{\mathrm{X}}^{k}\times\nonumber\\ &&\ket{(n-k)_{\mathrm{X}}}\bra{(n-k)_{\mathrm{X}}}\otimes\ket{k_{\mathrm{X'}}}\bra{k_{\mathrm{X'}}},\ea
followed by photon counting on the reflected mode $\mathrm{X'}$, and $R_{\mathrm{X}}$ is the reflectance of the beam splitter on mode X. The latter measurement is described by the two-outcome POVM $\{\Pi^{(0)}_{\mathrm{X'}},\Pi^{(1)}_{\mathrm{X'}}\}$ where
\ba
\Pi^{(0)}_{\mathrm{X'}}&=&\sum_{j\geq 0}(1-\eta_{\mathrm{X}})^j|j_{\mathrm{X'}}\rangle\langle j_{\mathrm{X'}}|,\\
\Pi^{(1)}_{\mathrm{X'}}&=&\sum_{j> 0}(1-(1-\eta_{\mathrm{X}})^j)\ket{j_{\mathrm{X'}}}\langle j_{\mathrm{X'}}|
\ea
describe the cases in which the photon counter did not, and did click, respectively, and $\eta_{\mathrm{X}}$ is the quantum efficiency of the counter at mode X. Both $R$ and  $\eta_\mathrm{X}, \mathrm{X}=\mathrm{A},\mathrm{B}$, are free parameters that describe the demon and whose values can be optimised. If one leaves $R_{\mathrm{A}}$ and $R_{\mathrm{B}}$ independent, the optimisation process generally returns the trivial case of $R_{\mathrm{A}}=1$ and $R_{\mathrm{B}}=0$. This would be the analog of a thermal demon that removes all the gas from one half of the container and lets the other half expand. To avoid this trivial situation, we shall set $R_{\mathrm{A}}=R_{\mathrm{B}}=R$.

For each of the four possible outcomes of the photon counting $C\equiv(c_{\mathrm{A}},c_{\mathrm{B}})\in\{(0,0),(0,1),(1,0),(1,1)\}$, we compute the probability $P_C$ that this outcome happens, as well as $ \bar{n}_{\mathrm{B}|C}-\bar{n}_{\mathrm{A}|C}$ on the conditional state left in the transmitted beams. Here, $\bar{n}_{\mathrm{X}|C}$ is the average photon number at mode X conditioned on the outcome of the photon counting $C$. If the latter average is negative, the polarity of the capacitor is switched. Thus, the figure of merit to be optimised over the parameters $(R,\eta_{\mathrm{A}},\eta_{\mathrm{B}})$ of the demon is
\ba
   \langle \Delta n\rangle=\sum_{C} (-1)^{s(C)}P_C\,\langle \Delta n\rangle_C&,
  \ea
  with  $\langle \Delta n\rangle_C =\bar{n}_{\mathrm{B}|C}-\bar{n}_{\mathrm{A}|C}$,
where $s(C)=1$ if polarity of the capacitor should be switched and $s(C)=0$ otherwise.


\section{Results}

\subsection{Two uncorrelated thermal states at the same temperature}
\label{ssoriginal}

The case of two uncorrelated thermal states with $\bar{n}_{\mathrm{A}}=\bar{n}_{\mathrm{B}}=\bar{n}$ is the one studied in Ref.~\cite{Vidrighin+5:16}. As discussed there, one finds that the demon helps with $\langle\Delta n\rangle_{\mathrm{max}}=(16/27)\bar{n}$. This is obtained using $s(1,0)=1$ and $s(C)=0$ for the other three cases. As already mentioned, this can be understood from the super-Poissonian statistics of the thermal state. Before proceeding to the study of other input states, it is important to make two remarks. 

\begin{figure}[t!]
\includegraphics[width=0.48\textwidth]{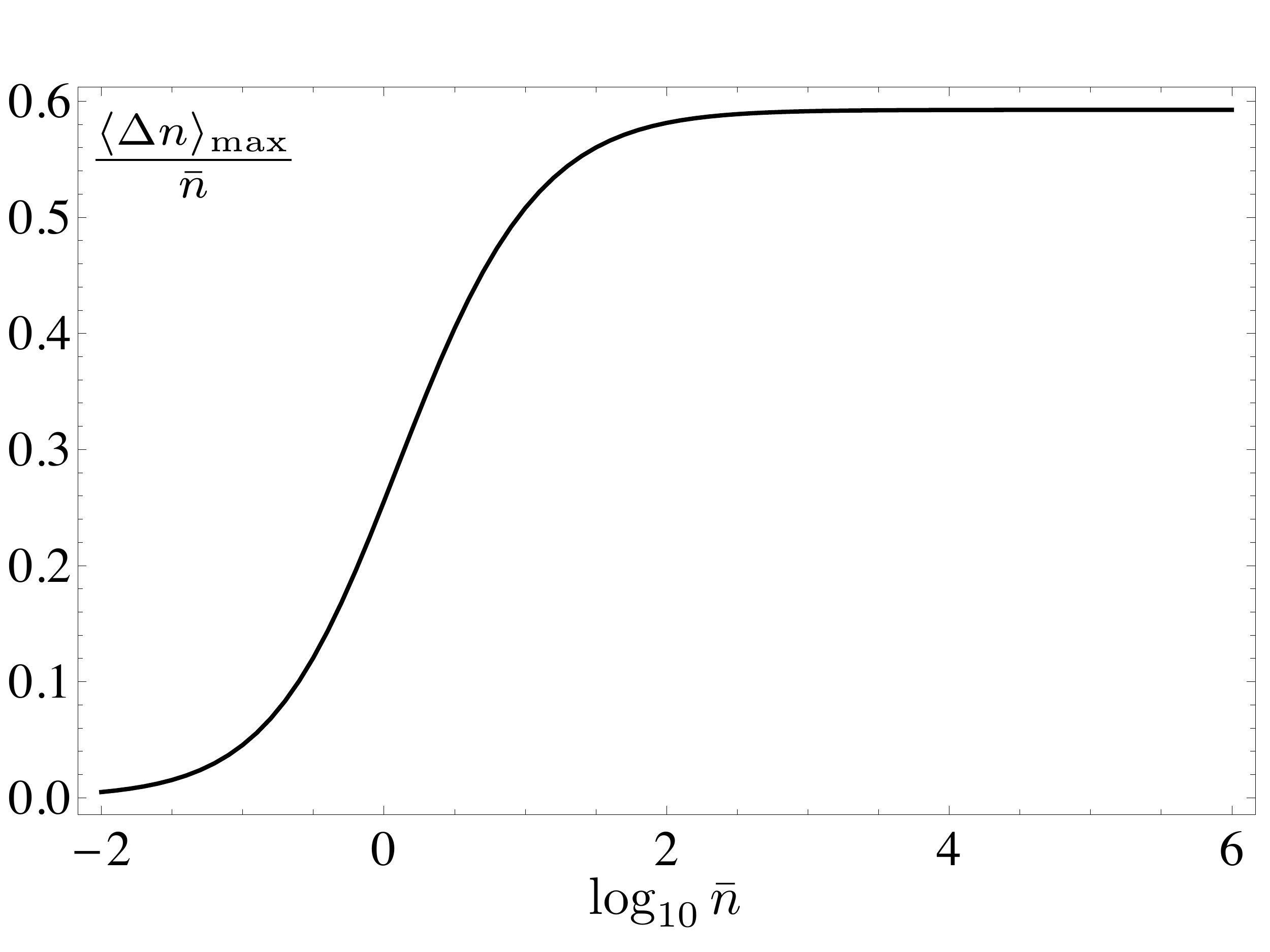}
 	\caption{Graph of maximum photon number difference that can be created by the Maxwell's demon when the average photon number in the two independent thermal state at the same temperature is varied. The plateau value is $16/27$, reached for $\bar{n}\approx 50$.}
	\label{fig:indepThermal}
\end{figure}

Firstly, the value of $(16/27)\bar{n}$ is only achievable for $\bar{n}\gg 1$ (which was the case in the experiment). This is implicit in their calculation -- arising from the fact that the optimisation was done over the probabilities of having a click in each mode [$p_{\mathrm{X}}=\mathrm{tr}\{(U\rho_{\mathrm{X}}U^\dagger)\Pi^{(1)}_{\mathrm{X}'}\}$ $(\mathrm{X}=\mathrm{A}, \mathrm{B})$], assuming that the whole range $0\leq p_{\mathrm{X}}\leq 1$ is accessible. Here $U\rho_\mathrm{X} U^\dagger$ is the state of $\rho_\mathrm{X}$ after the beam-splitter with unitary $U$ effecting the change described by Eq.~\eqref{beamsplitter}. However, using $\rho_\mathrm{X}=\rho^\mathrm{th}$, one finds that
\ba\label{clickprob}
p_{\mathrm{X}}=\frac{R\eta_{\mathrm{X}}\lambda}{1-\lambda+R\eta_{\mathrm{X}}\lambda}.
\ea Thus $
0\leq p_{\mathrm{X}}\leq \lambda$ for $0\leq R\eta_\mathrm{X}\leq 1$. In the limit $\bar{n}\rightarrow\infty$ one has $\lambda\rightarrow 1$ indeed. But, for example, if $\bar{n}=1$ one finds $p_{\mathrm{X}}\leq\frac{1}{2}$ and the optimal value $p_{\mathrm{A}}=2/3$, required to get $\langle\Delta n\rangle_{\mathrm{max}}=16/27$, would not be reached. In Fig.~\ref{fig:indepThermal}, we show the optimised  $\langle \Delta n\rangle_{\mathrm{max}}/\bar{n}$ as $\bar{n}$ is varied. The maximum value of $(16/27)\bar{n}$ is valid for $\bar{n}\gtrsim 50$. 

Secondly, a closer look shows that, in the $\bar{n}\gg 1$ limit, the optimal parameters are $R=\frac{2}{\bar{n}}$, $\eta_{\mathrm{A}}=1$ and $\eta_{\mathrm{B}}=\frac{1}{4}$. The fact that $\eta_{\mathrm{B}}\neq 1$ is slightly surprising: how can the demon gain any advantage by not detecting some photons after  they have been split out of the beam? As it turns out, this is not that strange. Consider the case $C=(0,0)$ when neither detector clicks. If $\eta_{\mathrm{A}}=\eta_{\mathrm{B}}=1$, then $\Delta n_{(0,0)}=0$ due to the obvious symmetry. By setting $\eta_\mathrm{B}<1$, the demon will guess that probably more photons were split into mode $\mathrm{B}'$ than into mode $\mathrm{A}'$, but was not detected due to the inefficient detector at beam B; thus, in agreement with the super-Poissonian statistics of the thermal state, it will guess that there are more photons left in B than in A. Calculation indeed shows that one can achieve $\Delta n_{(0,0)}>0$. As to why this actually helps in the total balance, and helps at best for $\eta_\mathrm{B}=\frac{1}{4}$, we are not able to provide an intuitive reason and have to rely on the optimisation. We highlighted this feature here because it will consistently recur in all the other examples that we are going to study.

\subsection{Two uncorrelated thermal states at different temperatures}

As a first case study, we consider a product of thermal states at different temperatures:
\ba
p(n_{\mathrm{A}},n_{\mathrm{B}})&=&(1-\lambda_{\mathrm{A}})\lambda_{\mathrm{A}}^{n_{\mathrm{A}}}\,(1-\lambda_{\mathrm{B}})\lambda_{\mathrm{B}}^{n_{\mathrm{B}}}.
\ea
Without loss of generality, we assume that the temperature of mode B is higher than that of mode A, so that $\bar{n}_{\mathrm{B}}>\bar{n}_{\mathrm{A}}$.

Obviously, with this state, the capacitor can be charged even without the demon. The average photon number difference created across the plates of the capacitor without the demon is simply $\langle\Delta n\rangle=\bar{n}_{\mathrm{B}}-\bar{n}_{\mathrm{A}}$. We would like to determine if the demon can still provide an advantage.

The calculation proceeds similarly as in Ref.~\cite{Vidrighin+5:16}, but we do not assume $\bar{n}_i\gg 1$ or $R\rightarrow 0$. The details are listed in Table~\ref{stbl:one} where we denoted by $p_{\mathrm{A}/\mathrm{B}}$ the probability that the counter on each mode clicks.

\begin{table}[t!]
\caption{\label{stbl:one}%
Uncorrelated thermal states at different temperatures. For simplicity of reading we have scaled $\langle\Delta n\rangle_C$ by a factor of $(1-R)P_{(0,0)}$ and denoted $\delta_1=R\bar{n}_{\mathrm{A}}\bar{n}_{\mathrm{B}}(\eta_{\mathrm{A}}-\eta_{\mathrm{B}})$, $\delta_2=R\bar{n}_{\mathrm{B}}[(\bar{n}_{\mathrm{B}}-\bar{n}_{\mathrm{A}})\eta_{\mathrm{B}}+\bar{n}_{\mathrm{A}}\eta_{\mathrm{A}}(2+R\bar{n}_{\mathrm{B}}\eta_{\mathrm{B}})]>0$, and $\delta_3=R\bar{n}_{\mathrm{A}}[(\bar{n}_{\mathrm{B}}-\bar{n}_{\mathrm{A}})\eta_{\mathrm{A}}-\bar{n}_{\mathrm{B}}\eta_{\mathrm{B}}(2+R\bar{n}_{\mathrm{A}}\eta_{\mathrm{A}})]$. In the table, $p_{\mathrm{X}}$, the probability of a click at mode X, is given by Eq.~\eqref{clickprob}.}
\centering
    \begin{tabular}{c@{\qquad}c@{\qquad}c}
    \hline \hline
   $C$ & $P_C$ & $\langle \Delta n\rangle_C/(P_{(0,0)}(1-R))$ \\  \hline
    (0,0) & $(1-p_{\mathrm{A}})(1-p_{\mathrm{B}})$  & $\bar{n}_{\mathrm{B}}-\bar{n}_{\mathrm{A}}+\delta_1$ \\ 
    (0,1) & $(1-p_{\mathrm{A}})p_{\mathrm{B}}$ & $2\bar{n}_{\mathrm{B}}-\bar{n}_{\mathrm{A}}+\delta_2$ \\
    (1,0) & $p_{\mathrm{A}}(1-p_{\mathrm{B}})$ & $\bar{n}_{\mathrm{B}}-2\bar{n}_{\mathrm{A}}+\delta_3$
    \\ (1,1) & $p_{\mathrm{A}}p_{\mathrm{B}}$  & $2\bar{n}_{\mathrm{B}}-2\bar{n}_{\mathrm{A}}+\delta_2+\delta_3-\delta_1$ \\ 
    \hline \hline
    \end{tabular}
\qquad
\end{table}

Now except the case $C=(0,1)$ where $\langle \Delta n\rangle_C$ is surely positive, the other three cases could potentially be negative depending on the values of the parameters. To find the optimal strategy of the demon, we numerically maximize $\langle\Delta n\rangle$ for all eight possible combinations of $s(C)$ for the other three cases for different range of values of $\bar{n}_{\mathrm{A}}$ and $\bar{n}_{\mathrm{B}}$. We first observed that for the situation of $\bar{n}_{\mathrm{B}}\gg\bar{n}_{\mathrm{A}}$, the best improvement by the demon is negligible, as it would be for the thermal demon.

\begin{figure}[b!]
\includegraphics[width=0.48\textwidth]{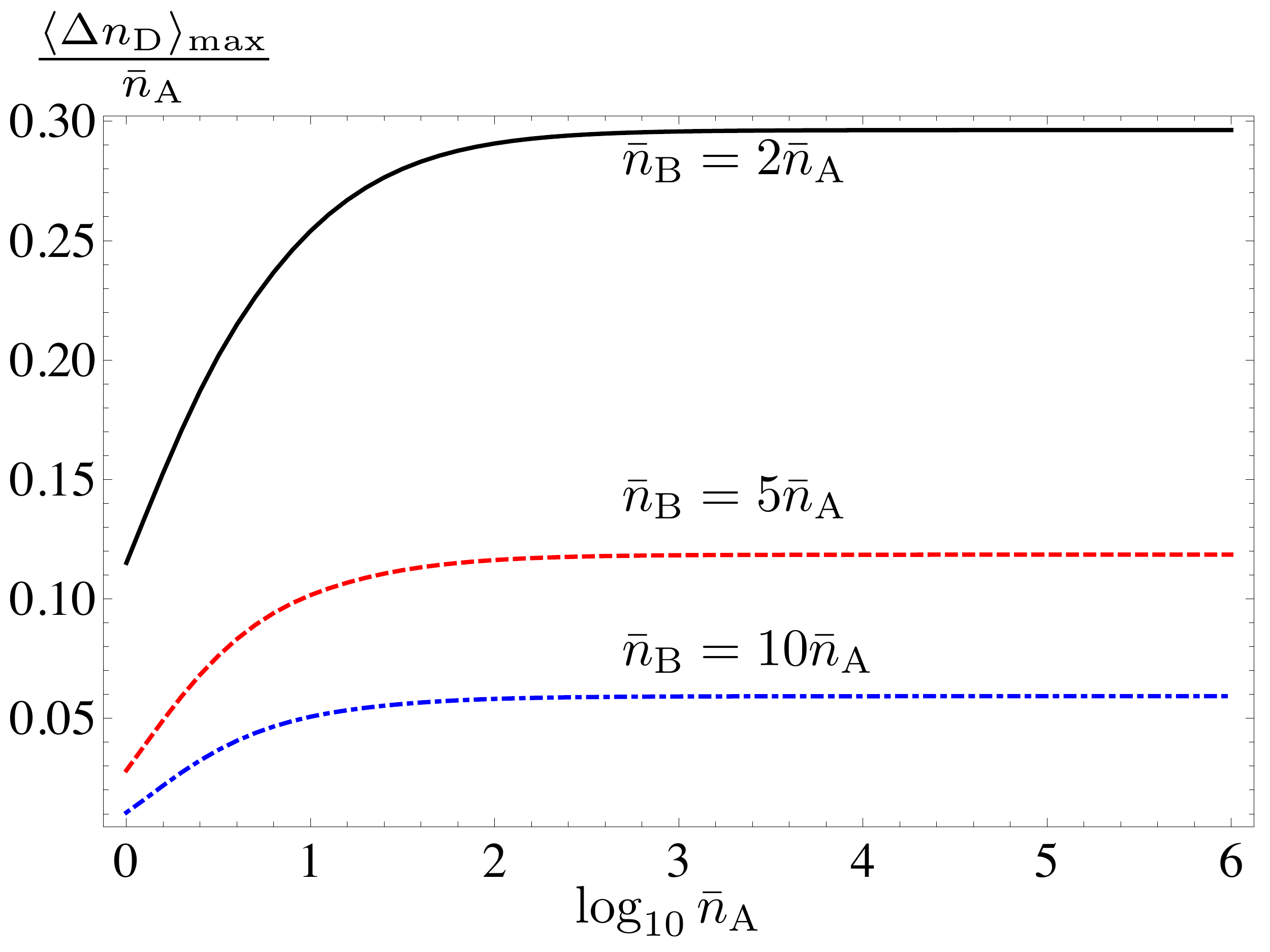}
 	\caption{(Color online) Uncorrelated Thermal States: Efficacy of Demon over different values of $\bar{n}_{\mathrm{A}}$ with fixed ratio of $\bar{n}_{\mathrm{B}}/\bar{n}_{\mathrm{A}}$. Here we consider the photon number difference gained due to the action of the demon only which is given by $\langle\Delta n_{\mathrm{D}}\rangle=\langle\Delta n\rangle-(\bar{n}_{\mathrm{B}}-\bar{n}_{\mathrm{A}})(1-R)$, since the latter subtracted quantity would be the photon number difference achievable even without the demon's action. The denominator of $\bar{n}_{\mathrm{A}}$ is merely there for scaling purposes. $\langle\Delta n_{\mathrm{D}}\rangle_{\mathrm{max}}$ is the maximum of $\langle\Delta n_{\mathrm{D}}\rangle$ achieved by the optimisation of the parameters.}
	\label{DiffTemp}
\end{figure}

Focusing then on the interesting regime where $\bar{n}_{\mathrm{B}}\gtrsim\bar{n}_{\mathrm{A}}$, we find that the maximum is always achieved for putting only $s(1,0)=1$. The average photon number difference created is then 
\begin{widetext}
\ba
\langle\Delta n\rangle=(1-R)\left[\bar{n}_{\mathrm{B}}-\bar{n}_{\mathrm{A}}+\frac{2\widetilde{R}_{\mathrm{A}}(\bar{n}_{\mathrm{A}}(1+\widetilde{R}_{\mathrm{B}})(2+\widetilde{R}_{\mathrm{A}})-\bar{n}_{\mathrm{B}}(1+\widetilde{R}_{\mathrm{A}}))}{(1+\widetilde{R}_{\mathrm{A}})^2(1+\widetilde{R}_{\mathrm{B}})^2}\right],
\ea
\end{widetext}
where $\widetilde{R}_\mathrm{X}\equiv\bar{n}_{\mathrm{X}}R\eta_{\mathrm{X}}$.
For an arbitrary $\bar{n}_{\mathrm{A}}$ and $\bar{n}_{\mathrm{B}}$, the maximum of this expression with the constraint that $0\leq \eta_\mathrm{X},R\leq 1$ is very lengthy and only numerical values can be obtained. For a better appreciation of how effective the Maxwell Demon is over a range of $\bar{n}$, we include a plot of the efficacy of Maxwell's Demon over $\bar{n}_{\mathrm{A}}$ in Fig.~\ref{DiffTemp} for several different fixed ratio of $\bar{n}_{\mathrm{B}}/\bar{n}_{\mathrm{A}}$. Notice that in this case of $\bar{n}_{\mathrm{B}}\neq \bar{n}_{\mathrm{A}}$, even without the demon, there would be a net average photon number difference. Hence, here we consider the photon number difference gained due to the action of the demon only which is given by $\langle\Delta n_{\mathrm{D}}\rangle=\langle\Delta n\rangle-(\bar{n}_{\mathrm{B}}-\bar{n}_{\mathrm{A}})(1-R)$, since the latter subtracted quantity would be the photon number difference achievable even without the demon's action.

\begin{figure}[t!]
\includegraphics[width=0.48\textwidth]{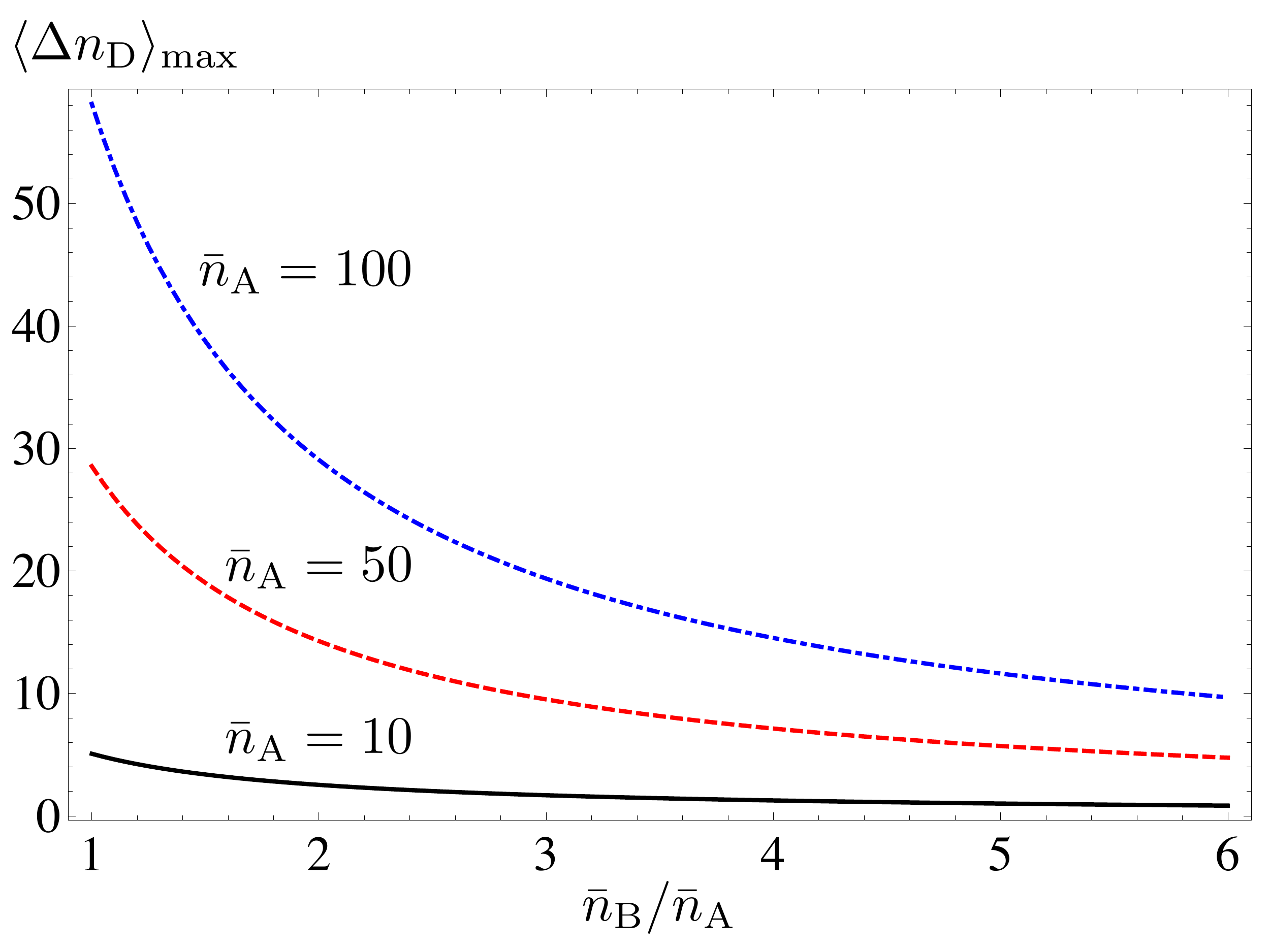}
 	\caption{(Color online) Uncorrelated Thermal States: Demon help is maximal when $\bar{n}_{\mathrm{A}}=\bar{n}_{\mathrm{B}}$, even at low $\bar{n}_{\mathrm{A}}$.}
	\label{ratio}
\end{figure}

As a remark, we note that for the case $\bar{n}_{\mathrm{A}}=\bar{n}_{\mathrm{B}}=1$, $\langle \Delta n\rangle_{\mathrm{max}}\approx 0.255<16/27$, achieved with $\eta_{\mathrm{B}}\approx 0.427$, $\eta_{\mathrm{A}}=1$, and $R\approx 0.344$.  Making connections with Ref.~\cite{Vidrighin+5:16}, in terms of the probabilities $p_{\mathrm{A}}$ and $p_{\mathrm{B}}$, one notes that $\widetilde{R}_\mathrm{X}=(1-p_{\mathrm{X}})^{-1}-1$, and finds the simpler expression
\ba
\langle\Delta n\rangle &=&(1-R)\big\{\bar{n}_{\mathrm{B}}-\bar{n}_{\mathrm{A}}+2p_{\mathrm{A}}(1-p_{\mathrm{B}})\nonumber\\&&\times[\bar{n}_{\mathrm{A}}(2-p_{\mathrm{A}})-\bar{n}_{\mathrm{B}}(1-p_{\mathrm{B}})]\big\}.
\ea
For the case where the average photon number in the two modes is quite large, the maximum value is approximately given by
\ba
\langle\Delta n\rangle_{\text{max}}\approx\bar{n}_{\mathrm{B}}-\bar{n}_{\mathrm{A}}+\frac{16}{27}\frac{\bar{n}_{\mathrm{A}}^2}{\bar{n}_{\mathrm{B}}},
\ea
obtained for 
$R=\frac{2}{\bar{n}_{\mathrm{A}}}$, $\eta_{\mathrm{A}}=1$, and $\eta_{\mathrm{B}}=\frac{3\bar{n}_{\mathrm{B}}-2\bar{n}_{\mathrm{A}}}{4\bar{n}_{\mathrm{B}}}$.
In terms of the probabilities, this translates to 
$p_{\mathrm{A}}=\frac{2}{3}$ and $p_{\mathrm{B}}=1-\frac{2}{3}\frac{\bar{n}_{\mathrm{A}}}{\bar{n}_{\mathrm{B}}}$. Thus the demon's action contributes an improvement of $\frac{16}{27}\frac{\bar{n}_{\mathrm{A}}^2}{\bar{n}_{\mathrm{B}}} \leq \frac{16}{27}\bar{n}_{\mathrm{A}}$, with equality if and only if $\bar{n}_{\mathrm{A}}=\bar{n}_{\mathrm{B}}=\bar{n}$. In other words, as expected, the demon helps, and its help is maximal when the two thermal states have the same temperature. This is true even without the constraints of large average photon numbers, as can be seen in the graph in Fig.~\ref{ratio}.

\begin{figure}[h!]
\includegraphics[width=0.48\textwidth]{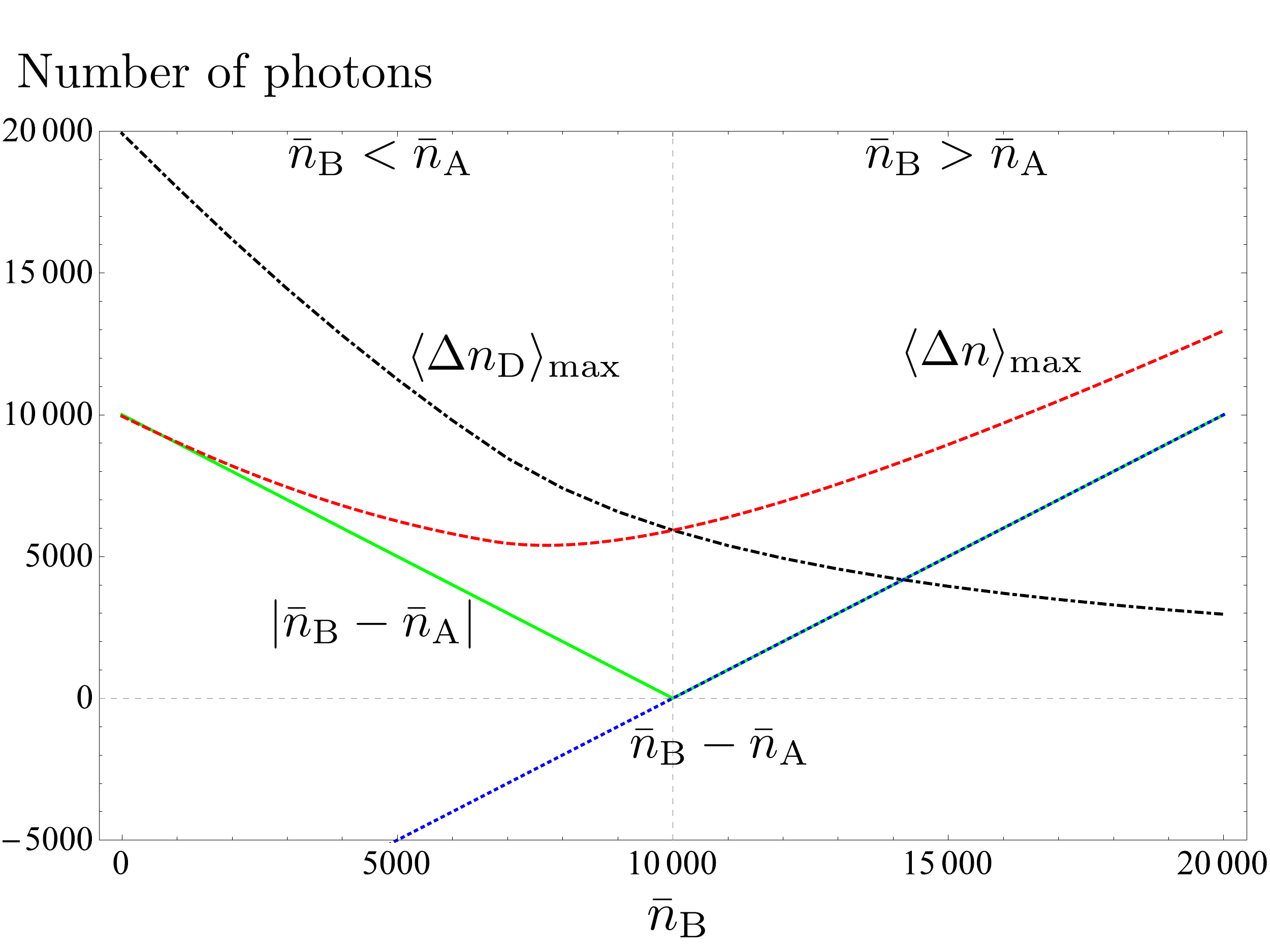}
 	\caption{(Colour online) Uncorrelated Thermal States: the blue dotted line is the protocol being implemented without the demon which is just the difference of $\bar{n}_{\mathrm{B}}-\bar{n}_{\mathrm{A}}$. Shown also for comparison is the green solid line which is the optimal protocol by changing the charge bias when $\bar{n}_{\mathrm{B}}<\bar{n}_{\mathrm{A}}$, but still without the action of the demon. The red dashed curve shows the maximum work that can be extracted with the demon while keeping the charging bias as if $\bar{n}_{\mathrm{B}}\geq\bar{n}_{\mathrm{A}}$. Notice that by doing so, even in the regime of $\bar{n}_{\mathrm{B}}<\bar{n}_{\mathrm{A}}$, the demon achieves $\langle\Delta n\rangle_{\mathrm{max}}>0$. This shows that the demon can create a backflow, overcoming an unfavorable bias. The plot is done by varying $\bar{n}_{\mathrm{B}}$ whilst keeping $\bar{n}_{\mathrm{A}}=10^4$. The black dash-dotted line shows the contribution of the demon only, which is the difference between the red dashed curve and the blue dotted line.}
	\label{backflow}
\end{figure}

At this juncture, the natural question one is led to is: Can the demon create backflows? That is, can the demon charge up the capacitor in a direction that is against thermodynamic directionality? This would be the analog of the original demon having to transfer the fast molecules into bath B starting from a bias $T_{\mathrm{A}}>T_{\mathrm{B}}$ --- which is surely possible but will cost the demon more effort. For the setup under study, we keep the demon's operation the same [i.e.~only $s(1,0)=1$] but allow $\bar{n}_{\mathrm{B}}<\bar{n}_{\mathrm{A}}$,and optimise over $\eta_{\mathrm{A}}$, $\eta_{\mathrm{B}}$ and $R$. The result is shown in Fig.~\ref{backflow}. Just like the original demon, ours can generate backflows. In the backflow region ($\bar{n}_{\mathrm{B}}<\bar{n}_{\mathrm{A}}$), the net contribution of the demon $\langle\Delta n_{\mathrm{D}}\rangle_{\mathrm{max}}$ (the black dash-dotted line) is much higher, as expected. However, this does not imply that it is better to operate the system in the backflow region, which would be strange: as seen in Fig.~\ref{backflow}, for a fixed value of $|\bar{n}_{\mathrm{B}}- \bar{n}_{\mathrm{A}}|$, the value of $\langle\Delta n\rangle_{\mathrm{max}}$ (the red dashed line) is larger in the region of direct flow, once again as expected.

\subsection{Split thermal state}

Next we consider the state obtained by sending a thermal state through a beam splitter: $\rho_{\mathrm{AB}}=U\rho^{\mathrm{th}}(\beta')U^\dagger$, where $U=\mathrm{exp}\big[\theta(a^\dagger b\mathrm{e}^{\mathrm{i}\phi}-a b^\dagger\mathrm{e}^{-\mathrm{i}\phi})\big]$ and the reflectance of the beam splitter is given by $R=\sin^2\theta$ \cite{Campos+2:89}. The photon number distribution is now given by \cite{Loudon:00}
\ba
p(n_{\mathrm{A}},n_{\mathrm{B}})&=&\frac{1}{1+\bar{n}_{\mathrm{in}}}\frac{(n_{\mathrm{A}}+n_{\mathrm{B}})!}{n_{\mathrm{A}}!n_{\mathrm{B}}!}\nonumber\\
&&\times\left(\frac{\bar{n}_{\mathrm{A}}}{1+\bar{n}_{\mathrm{in}}}\right)^{n_{\mathrm{A}}}\left(\frac{\bar{n}_{\mathrm{B}}}{1+\bar{n}_{\mathrm{in}}}\right)^{n_{\mathrm{B}}},
\ea
where $\bar{n}_{\mathrm{in}}=1/(\mathrm{e}^{\beta'\hbar\omega}-1)$. Using $\sum_{k=0}^\infty{k+a \choose k} x^k=(1-x)^{-(1+a)}$, we see that the marginal states are still thermal with average number $\bar{n}_{\mathrm{A}}=\bar{n}_{\mathrm{in}}\cos^2\theta$ and $\bar{n}_{\mathrm{B}}=\bar{n}_{\mathrm{in}}\sin^2\theta$, but now there are correlations (in particular, it holds $\langle n_\text{A}n_\text{B}\rangle=2\bar{n}_\text{A}\bar{n}_\text{B}$).

The result of the calculation is presented in Table \ref{stbl:two}. The expressions are unpleasant but one feature is clear: all the $\langle\Delta n\rangle_C$ are proportional to $\bar{n}_{\mathrm{B}}-\bar{n}_{\mathrm{A}}$ with a positive factor. In other words, because of the correlations created by the beam-splitter, the sign of the difference of photon numbers is not modified for any value of $C$. The demon cannot help when such correlations are present and one checks that $\langle \Delta n\rangle=(1-R)(\bar{n}_{\mathrm{B}}-\bar{n}_{\mathrm{A}})$.

\begin{table}
\caption{\label{stbl:two}%
Split thermal state. We denote $\widetilde{R}_{\mathrm{X}}\equiv\bar{n}_{\mathrm{X}} R\eta_{\mathrm{X}}$. To facilitate reading, we have scaled the expressions by $P_{(0,0)}=1/(1+\widetilde{R}_{\mathrm{A}}+\widetilde{R}_{\mathrm{B}})$ and denoted $K\equiv 1+1/P_{(0,0)}= 2+\widetilde{R}_{\mathrm{A}}+\widetilde{R}_{\mathrm{B}}$. The expression for $\langle\Delta n\rangle_C$ is also scaled by $1-R$. In the table, $K'=2(K-1)[3+\widetilde{R}_{\mathrm{A}}\widetilde{R}_{\mathrm{B}}+(\widetilde{R}_{\mathrm{A}}+\widetilde{R}_{\mathrm{B}})(\widetilde{R}_{\mathrm{A}}+\widetilde{R}_{\mathrm{B}}+3)]+\widetilde{R}_{\mathrm{A}} \widetilde{R}_{\mathrm{B}}(\widetilde{R}_{\mathrm{A}}+\widetilde{R}_{\mathrm{B}})^2$. }
\centering
    \begin{tabular}{c@{\qquad}c@{\qquad}c}
    \hline \hline
   $C$ & $P_C/P_{(0,0)}$ & $\langle \Delta n\rangle_C/(P_{(0,0)}(1-R))$ \\  \hline
    (0,0) & 1  & $(\bar{n}_{\mathrm{B}}-\bar{n}_{\mathrm{A}}$) \\ 
    (0,1) & $\frac{\widetilde{R}_{\mathrm{B}}}{(1+\widetilde{R}_{\mathrm{A}})}$  & $\frac{(\bar{n}_{\mathrm{B}}-\bar{n}_{\mathrm{A}})(K+\widetilde{R}_{\mathrm{A}})}{(1+\widetilde{R}_{\mathrm{A}})}$ \\
    (1,0) & $\frac{\widetilde{R}_{\mathrm{A}}}{(1+\widetilde{R}_{\mathrm{B}})}$ & $\frac{(\bar{n}_{\mathrm{B}}-\bar{n}_{\mathrm{A}})(K+\widetilde{R}_{\mathrm{B}})}{(1+\widetilde{R}_{\mathrm{B}})}$\\  
    (1,1) & $\frac{\widetilde{R}_{\mathrm{A}}\widetilde{R}_{\mathrm{B}} K}{(1+\widetilde{R}_{\mathrm{A}})(1+\widetilde{R}_{\mathrm{B}})}$  & $\frac{(\bar{n}_{\mathrm{B}}-\bar{n}_{\mathrm{A}})K'}{(1+\widetilde{R}_{\mathrm{A}})(1+\widetilde{R}_{\mathrm{B}})K}$\\
    \hline \hline
    \end{tabular}
\qquad
\end{table}

\subsection{Number-correlated state (two-mode squeezed state)}\label{paratmss}

Next we consider the number correlated state
\ba
p(n_{\mathrm{A}},n_{\mathrm{B}})&=&\frac{1}{1+\bar{n}}\left(\frac{\bar{n}}{1+\bar{n}}\right)^{n_{\mathrm{A}}}\,\delta_{n_{\mathrm{A}},n_{\mathrm{B}}}\,.
\ea
This distribution is that of the two-mode squeezed state
\begin{equation}
    |\psi\rangle_{\mathrm{AB}}=\frac{1}{\cosh r}\sum_{n=0}^\infty (\tanh r)^n |nn \rangle,
    \end{equation}
where $r$ is the squeezing parameter and $\bar{n}=\sinh^2 r$. For this state, the calculations are heavy and the final optimisation must be done numerically; so we give them in Appendix \ref{apptmss}. We include here in Fig.~\ref{TMSSgraph} the Demon's efficiency over $\bar{n}$.

\begin{figure}[b!]
\includegraphics[width=0.48\textwidth]{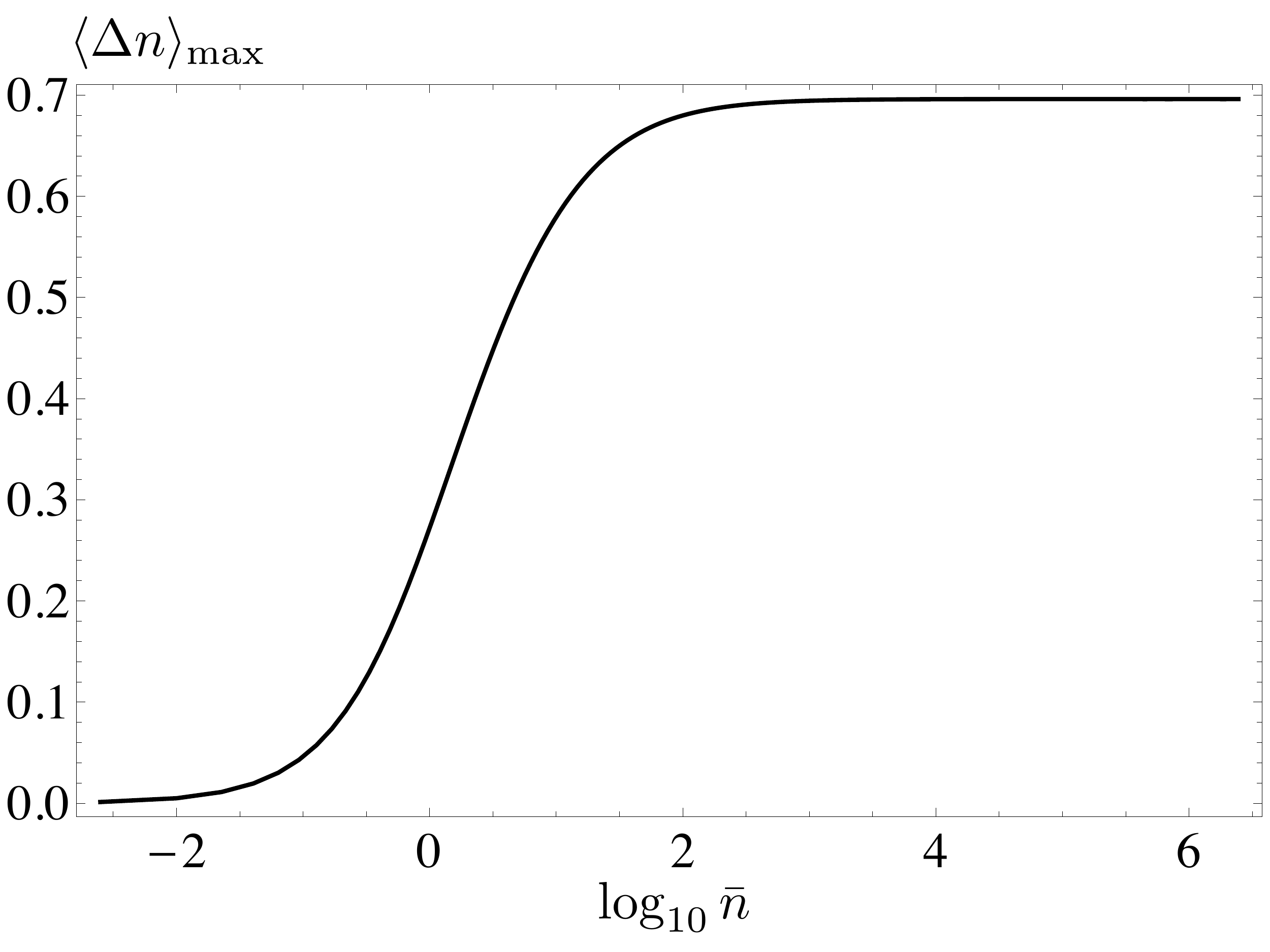}
  	\caption{Graph of maximum photon number difference that can be created by the Maxwell's demon when the average photon number in the number-correlated state is varied.}
	\label{TMSSgraph}
\end{figure}

The overall result is that the demon helps but very little, with
\ba
\langle\Delta n\rangle_{\mathrm{max}}\approx 0.7,\,\,&\textrm{  for  }&\, \bar{n}\longrightarrow\infty.
\ea
It is instructive to give a qualitative understanding of this result. Recall that the demon basically performs photon subtraction. Upon detecting a photon in (say) the reflection of mode A, as before the demon knows that probably there are several photons left in that mode, because of the marginal thermal statistics. But now it also knows that in the other mode there are many photons too: to be precise, there is \textit{one more photon}. So, now the demon has to set $s(0,1)=1$, instead of $s(1,0)=1$ as for the uncorrelated thermal states; besides, the difference in photon numbers will never be larger than 1 on average. Finally, as comparison to the other states, we note that for $\bar{n}=1$, $\langle \Delta n\rangle_{\mathrm{max}}\approx 0.272$, which is achieved with $R\approx 0.373$, $\eta_{\mathrm{A}}\approx 0.415$, and $\eta_{\mathrm{B}}=1$. The role of the efficiency for the two detectors are swapped here as compared to the super-Poissonian case, in keeping with what we expect. Indeed, by setting $\eta_{\mathrm{A}}<1$ the demon guesses that there were probably more photons in mode $\mathrm{A}'$ than $\mathrm{B}'$ in case of no-detection. Due to the strict photon number correlation, it infers that there are probably fewer photons in mode A than in mode B.

\subsection{Number-anticorrelated states}\label{NOON}

The previous result immediately evokes its counterpart: we expect the demon to be  very efficient if the modes are \textit{anti-correlated} in numbers, for instance, a mixture of $\ket{n_{\mathrm{A}},0_{\mathrm{B}}}$ and $\ket{0_{\mathrm{A}},n_{\mathrm{B}}}$ should increase the efficiency of the demon. We are therefore going to study states with statistics
\ba\label{puncorr}
p(0,0)=q_0\,&,&\;p(n,0)=p(0,n)=q_n/2,\,\text{for} \,\,n>0,
\ea and all the other $p(n_{\mathrm{A}},n_{\mathrm{B}})=0$. The normalisation condition is $\sum_n q_n=1$. For all these states, $P_C=0$ for $C=(1,1)$; and, as soon as one counter clicks, the demon knows that that mode contains all the remaining photons.

Let us first consider the simple case $q_k=\delta_{k,m}$ with $\bar{n}_{\mathrm{A}}=\bar{n}_{\mathrm{B}}=m/2$. When the second counter clicks [$C=(0,1)$], the value of $P_C\langle \Delta n\rangle_C$ is
\ba
&&\sum_{k=0}^m \binom{m}{k}R^k(1-R)^{m-k}(1-(1-\eta_{B})^k)(m-k)\nonumber\\=&&\frac{m(1-R)}{2}\left[1-(1-R\eta_{\mathrm{B}})^{m-1}\right]\,.
\label{anticor01}
\ea 

To obtain the expression in the case where only the first counter clicks [$C=(1,0)$], just add a negative sign in Eq.~\eqref{anticor01} and swap out the $\eta_{B}$ with $\eta_{A}$.

When neither counter clicks, if $\eta_\mathrm{B}\leq\eta_\mathrm{A}$ one can still have an estimate $\langle \Delta n\rangle_{(0,0)}\geq 0$: indeed, for $C=(0,0)$,
\ba
P_C\langle \Delta n\rangle_C&=&\frac{m(1-R)}{2} \nonumber\\
&&\times\left[(1-R\eta_{\mathrm{B}})^{m-1}-(1-R\eta_{\mathrm{A}})^{m-1}\right]\!.
\ea 
All in all, by setting $s(1,0)=1$, the demon achieves
\ba
\langle\Delta n\rangle=&m&(1-R)[1-(1-R\eta_{\mathrm{A}})^{m-1}],
\ea with $\eta_\mathrm{A}\geq\eta_\mathrm{B}$. For $m=0,1$, this yields 0 as it should. Maximising over the demon's parameters, one finds
\ba 
\langle\Delta n\rangle_{\mathrm{max}}&=&(m-1)\,m^{\frac{1}{1-m}}\nonumber\\
&\approx&m-1-\ln m,\,\,\mathrm{for}\,\,m\gg 1,
\ea
achieved for $\eta_{\mathrm{A}}=1$ and $R=1-m^{\frac{1}{1-m}}$. Recalling that $m=2\bar{n}$, on this state the demon comes close to the absolute maximal performance, as expected.

After this simple example, let us study number-anticorrelated states whose marginals are thermal. From Eq.~\eqref{puncorr}, we have $p(n_{\mathrm{A}}=n)=p(n_{\mathrm{B}}=n)=q_{n}/2$ for $n>0$, and therefore we want to impose $q_n=\frac{2}{1+\bar{n}}\left(\frac{\bar{n}}{1+\bar{n}}\right)^{n}$. This implies $\sum_{n>0}q_n=2(1-1/(1+\bar{n}))$. Checking that this sum is smaller than one constrains the mean photon number in each mode to satisfy $\bar{n}\leq 1$.

Let us finish the calculation for the case $\bar{n}=1$, corresponding to $q_0=0$ and $q_{n>0}=2^{-n}$. With the results obtained above, it is straightforward to work out the mean photon number difference
\begin{equation}
    \langle\Delta n\rangle=\frac{2(1-R)R\eta_{\mathrm{A}}(2+R \eta_{\mathrm{A}})}{(1+R\eta_{\mathrm{A}})^2}.
\end{equation}
The function is maximised for $\eta_{\mathrm{A}}=1$ and for $R$ solution of the equation $R^3+3R^2+4R-2=0$, i.e. $R\approx 0.379$. The corresponding value for the maximum photon number  difference is then the solution of the equation $4x^3-49x^2+272x-144=0$, that is
\ba \langle\Delta n\rangle_{\mathrm{max}}&\approx & 0.589\,.\ea

\subsection{Quick overview}

\begin{figure}[t!]
\includegraphics[width=0.48\textwidth]{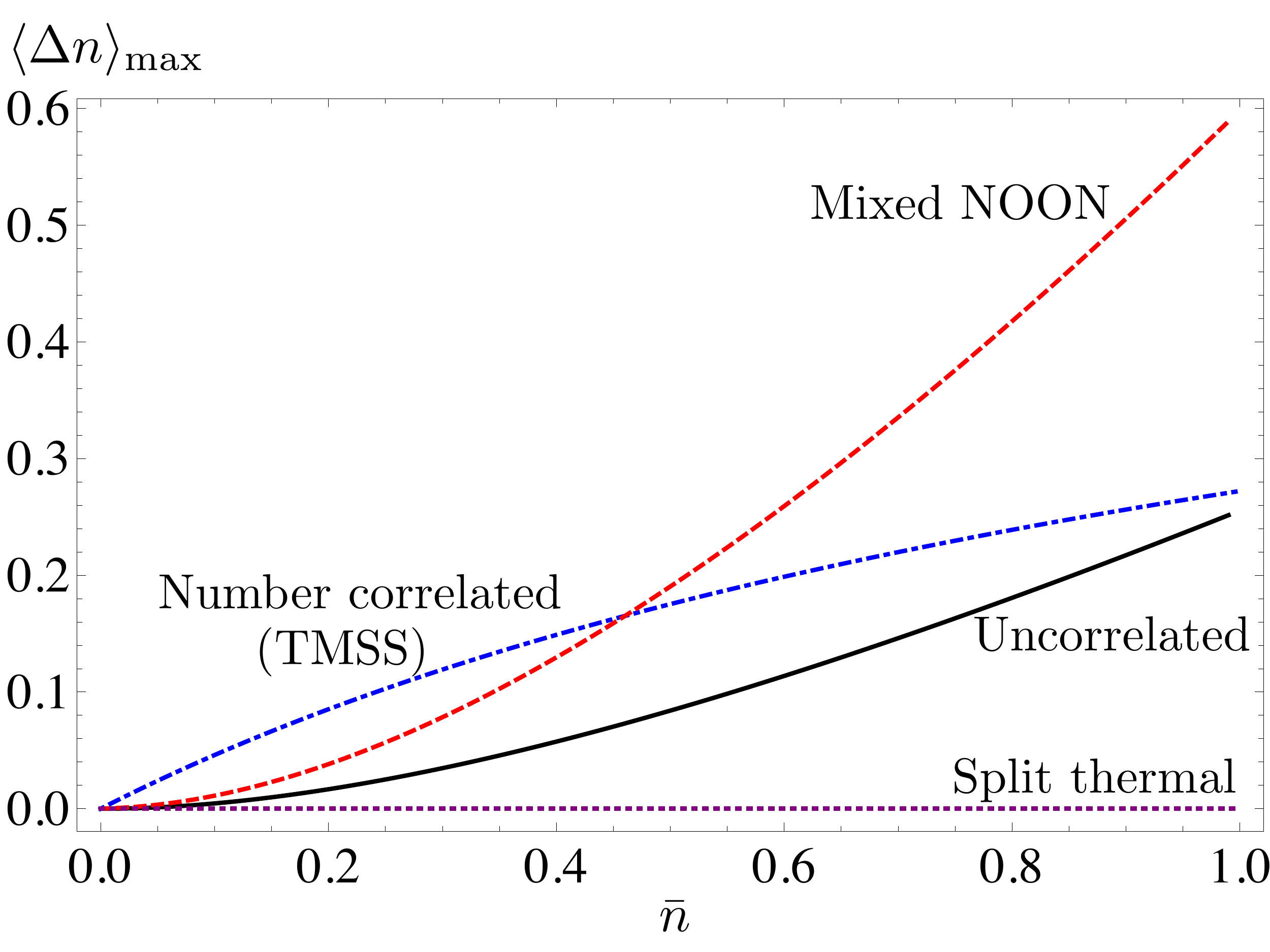}
 	\caption{(Colour online) Summary of the effect of the Demon over all the cases with thermal marginals. We kept to the limit $0\leq\bar{n}\leq 1$ because the mixed NOON states are only valid in that regime (Refer to section \ref{NOON} for details). For ease of viewing, we only restricted ourselves to $\bar{n}_{B}=\bar{n}_{A}$ in this graph, therefore $\langle\Delta n \rangle_{\text{max}}$ here is solely from the contribution of the demon. It is clear from this graph that both the quantum correlated states (TMSS and Mixed NOON) fare better than the classical (both uncorrelated and classically correlated) states under low $\bar{n}$. In the large $\bar{n}$ regime however (not shown in this graph), the classical (uncorrelated) states fare the best if we require the marginals to remain thermal.}
	\label{summary}
\end{figure}

Table \ref{ResultsTable} summarizes some of the results, namely the maximal contribution of the demon to the states that we studied, in the case $\bar{n}_{\mathrm{A}}=\bar{n}_{\mathrm{B}}$. The number-anticorrelated states empower the demon tremendously, the number correlated ones hinder its action. Fig.~\ref{summary} plots on one graph the comparison of the power of the demon for all the different sources as a function of $\bar{n}$, assuming $\bar{n}<1$, to comply with the constraint on the mixed NOON state. Thus, the optical demon behaves in good analogy with the thermal demon (Fig.~\ref{fig:intuitive}) even in this respect. Among the states with thermal marginals, in the limit of large $\bar{n}$, the original uncorrelated state seems to be the best option, insofar as anti-correlated states are not available.

It is also a recurrent feature that the optimisation leads to one of the $\eta$s being unity and the other less than unity (see discussion at the end of section \ref{ssoriginal} and \ref{paratmss}).

\begin{table}
		\centering
		\caption{Summary: the contribution of the demon for the states studied in this section, with $\bar{n}_{\mathrm{A}}= \bar{n}_{\mathrm{B}}=\bar{n}$. The first four states have thermal marginals.}\label{ResultsTable}
		\begin{tabular}{|c|c|c|}
		\hline
		 \multirow{2}{*}{State}	 &  $\langle\Delta n\rangle_{\mathrm{max}}$ & $\langle\Delta n\rangle_{\mathrm{max}}$ \\
		 &($\bar{n}=1$)&(Large $\bar{n}$)\\\hline
		 	Uncorrelated &$\approx 0.255$&$(16/27)\bar{n}$ \,\cite{Vidrighin+5:16}\\
		 	Split thermal state & 0&0\\
		 	Number-correlated &$\approx 0.272$&$\approx 0.7$\\
		 	Number-anticorrelated & $\approx 0.589$& -- \\
		 	Number-anticorrelated ($m$) & 0.5& $\approx 2\bar{n}$ \\\hline
		\end{tabular}	
	\end{table}

\section{The demon and single-copy passive states}

A quantum state is called \textit{passive} if it cannot be processed to extract work. Which states are actually passive depend on the rules of work extraction. Usually one considers rather abstract rules, allowing for very general operations: in this context, a famous result is that the thermal state is the only completely passive state, that is, the state that remains passive no matter how many copies of it are made available (see \cite{Skrzypczyk+2:15} and references therein). These definitions are at the basis of the resource theory of thermal operations that has been developed recently \cite{Brandao+4:13, Horodecki+Oppenheim:13,Goold+4:16}.

For our study, we are not going to rely on such general results, but an analysis of our various states and of the power of the demon in terms of resources may nevertheless be attempted. After all, the choice of thermal state in the original paper was motivated by its passivity; and the demon is definitely a resource, so it may be interesting to compare it to other resources.

We are going to work with a very restricted set of rules for work extraction, which is basically a rephrasing of the scenario studied above. Work is extracted by charging the capacitor in the scheme under consideration. Independent copies of the state are sent sequentially into the setup, so that only single-copy work extraction is considered. When the states are not thermal, we neglect the cost of preparing those states. As another free resource, we possess a thermal bath at temperature $T$. This bath can be coupled to each of the optical modes independently, resulting in the thermalisation of the partial state: in particular, the mean photon number becomes $\bar{n}_T$ after thermalisation \cite{Nagaj+3:02}.

With these rules, all and only those states such that $\bar{n}_{\mathrm{A}}=\bar{n}_{\mathrm{B}}=\bar{n}_T$ are passive. Indeed, if a state has $\bar{n}_{\mathrm{A}}\neq\bar{n}_{\mathrm{B}}$, a net charge can be created. And if a state has $\bar{n}_{\mathrm{A}}=\bar{n}_{\mathrm{B}}\neq\bar{n}_T$, one can thermalise one of the modes for free, thus creating $\bar{n}_{\mathrm{A}}\neq\bar{n}_{\mathrm{B}}$.

With respect to this classification, the uncorrelated thermal state of the original study is still passive, as it should. The split thermal state is not passive if the free thermal bath is naturally assumed to coincide with the bath that prepares that state. Indeed, after splitting one has  $\bar{n}_{\mathrm{A}}=\bar{n}_{\mathrm{B}}=\bar{n}_T/2$, so one can thermalise mode B to reach $\bar{n}_{\mathrm{B}}=2\bar{n}_{\mathrm{A}}$. Interestingly, recall that for this state the demon as implemented doesn't help at all. Thus, as a resource, \textit{the demon is incommensurable with state preparation}. Turning to the other states (number-correlated and -anticorrelated), their definition does not involve any thermal bath; so the choice of a $T$ for the bath is arbitrary given the parameters of the state. Unsurprisingly, we find that all states in these families are resources, insofar as one does not choose $T$ to match exactly $\bar{n}_{\mathrm{A}}=\bar{n}_{\mathrm{B}}=\bar{n}_T$.

\section{Conclusion}

We have further expanded the study of the optical simulation of the Maxwell demon demonstrated in Ref.~\cite{Vidrighin+5:16} to include the effects of correlation. Even without this study, ones knows that the information that the demon collects must be useful in order for it to extract work. In the original study, it was useful because it could signal a change in statistics in one and only one of the modes. We see, in this paper, that the usefulness decreases significantly if the numbers are positively correlated across the modes, and increases if they are negatively correlated. This is the same behavior a thermal demon would exhibit in the presence of similar correlations in the speed of the atoms between the two reservoirs.

On the one hand, there are some quantum elements in the demon under study: notably, it uses photon-counting, which cannot be described in a classical theory of light. Also, one may say that number anti-correlated states of optical fields ($g^{(2)}<1$) are necessarily quantum, and this demon's power is enhanced by them. On the other hand, though, all the statistics we studied could be simulated with classical systems: quantum correlations do not play any role for this demon. A practical implementation of a properly quantum demon, that would allow studying behaviors like those predicted in some information-theoretical papers \cite{Delrio+4:11,Braga+3:14, Chapman+Miyake:15}, is still lacking.

\begin{acknowledgements}
We acknowledge clarifying discussions with the authors of \cite{Vidrighin+5:16}, in particular Marco Barbieri, Vlatko Vedral and Mihai-Dorian Vidrighin; as well as with Tim Ralph, Paul Skrzypczyk, Christian Kurtsiefer and Stefan Nimmrichter. We also thank an anonymous referee for bringing up several points for improvement.
This research is supported by the National Research Foundation (NRF) Singapore under its Competitive Research Programme (CRP Award No. NRF-CRP12-2013-03), and NRF Singapore and the Ministry of Education, Singapore under the Research Centres of Excellence programme.
\end{acknowledgements}

\begin{widetext}
\appendix
\numberwithin{equation}{section}

\section{Details for the number-correlated state}\label{apptmss}

In this Appendix, we give details on the calculations for the number-correlated state presented in Section \ref{paratmss}.
We use a slightly different notation from the main text here: $\widetilde{R}_{\mathrm{X}}=\eta_{\mathrm{X}}R$. With this, the probabilities for the counting patterns of the demon are:
\ba
P_{(0,0)}&=&\frac{1}{1+\bar{n}(\widetilde{R}_{\mathrm{A}}+\widetilde{R}_{\mathrm{B}}-\widetilde{R}_{\mathrm{A}}\widetilde{R}_{\mathrm{B}})},
\\
P_{(0,1)}&=&\frac{\bar{n}\widetilde{R}_{\mathrm{B}}(1-\widetilde{R}_{\mathrm{A}})}{(1+\bar{n}\widetilde{R}_{\mathrm{A}})[1+\bar{n}(\widetilde{R}_{\mathrm{A}}+\widetilde{R}_{\mathrm{B}}-\widetilde{R}_{\mathrm{A}}\widetilde{R}_{\mathrm{B}})]},
\\
P_{(1,0)}&=&\frac{\bar{n}\widetilde{R}_{\mathrm{A}}(1-\widetilde{R}_{\mathrm{B}})}{(1+\bar{n}\widetilde{R}_{\mathrm{B}})[1+\bar{n}(\widetilde{R}_{\mathrm{A}}+\widetilde{R}_{\mathrm{B}}-\widetilde{R}_{\mathrm{A}}\widetilde{R}_{\mathrm{B}})]},
\\
P_{(1,1)}&=&\frac{\bar{n}\widetilde{R}_{\mathrm{A}}\widetilde{R}_{\mathrm{B}}[1+2\bar{n}+\bar{n}^2(\widetilde{R}_{\mathrm{A}}+\widetilde{R}_{\mathrm{B}}-\widetilde{R}_{\mathrm{A}}\widetilde{R}_{\mathrm{B}})]}{(1+\bar{n}\widetilde{R}_{\mathrm{A}})(1+\bar{n}\widetilde{R}_{\mathrm{B}})[1+\bar{n}(\widetilde{R}_{\mathrm{A}}+\widetilde{R}_{\mathrm{B}}-\widetilde{R}_{\mathrm{A}}\widetilde{R}_{\mathrm{B}})]}.
\ea

The mean photon number difference conditioned on each case is given by
\begin{equation}
\langle \Delta n\rangle_{(0,0)}=\frac{\bar{n}(1-R)(\widetilde{R}_{\mathrm{B}}-\widetilde{R}_{\mathrm{A}})}{1+\bar{n}(\widetilde{R}_{\mathrm{A}}+\widetilde{R}_{\mathrm{B}}-\widetilde{R}_{\mathrm{A}}\widetilde{R}_{\mathrm{B}})},
\end{equation}

\begin{equation}
\langle \Delta n\rangle_{(0,1)}=-\frac{(1-R)\left(1+\bar{n}\widetilde{R}_{\mathrm{A}}\left[4-2\widetilde{R}_{\mathrm{A}}+\bar{n}\left(\widetilde{R}_{\mathrm{A}}(3-2\widetilde{R}_{\mathrm{A}})+\widetilde{R}_{\mathrm{B}}(1-\widetilde{R}_{\mathrm{A}})^2\right)\right]\right)}{(1-\widetilde{R}_{\mathrm{A}})(1+\bar{n}\widetilde{R}_{\mathrm{A}})[1+\bar{n}(\widetilde{R}_{\mathrm{A}}+\widetilde{R}_{\mathrm{B}}-\widetilde{R}_{\mathrm{A}}\widetilde{R}_{\mathrm{B}})]}<0,
\end{equation}

\begin{equation}
\langle \Delta n\rangle_{(1,0)}=\frac{(1-R)\left(1+\bar{n}\widetilde{R}_{\mathrm{B}}\left[4-2\widetilde{R}_{\mathrm{B}}+\bar{n}\left(\widetilde{R}_{\mathrm{B}}(3-2
\widetilde{R}_{\mathrm{B}})+\widetilde{R}_{\mathrm{A}}(1-\widetilde{R}_{\mathrm{B}})^2\right)\right]\right)}{(1-\widetilde{R}_{\mathrm{B}})(1+\bar{n}\widetilde{R}_{\mathrm{B}})[1+\bar{n}(\widetilde{R}_{\mathrm{A}}+\widetilde{R}_{\mathrm{B}}-\widetilde{R}_{\mathrm{A}}\widetilde{R}_{\mathrm{B}})]}>0,
\end{equation}
and
\ba
\langle \Delta n\rangle_{(1,1)}&=&\frac{\bar{n}(1-R)(\widetilde{R}_{\mathrm{A}}-\widetilde{R}_{\mathrm{B}})(2+\bar{n}f_1+\bar{n}^2f_2+\bar{n}^3f_3)}
{(1+\bar{n}\widetilde{R}_{\mathrm{A}})(1+\bar{n}\widetilde{R}_{\mathrm{B}})[1+\bar{n}(\widetilde{R}_{\mathrm{A}}+\widetilde{R}_{\mathrm{B}}-\widetilde{R}_{\mathrm{A}}\widetilde{R}_{\mathrm{B}})][1+2\bar{n}+\bar{n}^2(\widetilde{R}_{\mathrm{A}}+\widetilde{R}_{\mathrm{B}}-\widetilde{R}_{\mathrm{A}}\widetilde{R}_{\mathrm{B}})]},
\ea
with
\ba
f_1&=&3+2\widetilde{R}_{\mathrm{A}}+2\widetilde{R}_{\mathrm{B}}-\widetilde{R}_{\mathrm{A}}\widetilde{R}_{\mathrm{B}},\\
f_2&=&2(2\widetilde{R}_{\mathrm{A}}+2\widetilde{R}_{\mathrm{B}}-\widetilde{R}_{\mathrm{A}}\widetilde{R}_{\mathrm{B}}),\\
f_3&=&\widetilde{R}_{\mathrm{B}}^2(1-\widetilde{R}_{\mathrm{A}})^2+\widetilde{R}_{\mathrm{A}}\widetilde{R}_{\mathrm{B}}(3-2\widetilde{R}_{\mathrm{A}})+\widetilde{R}_{\mathrm{A}}^2.
\ea
The fact that the expression for $\langle \Delta n\rangle_{(0,1)}$ is always negative and $\langle \Delta n\rangle_{(1,0)}$ always positive reflects that the demon knows, upon detection of a photon in one mode, that there is probably one more photon in the other mode. Hence, one should set $s(0,1)=1$, and $s(1,0)=0$ since, by our definition $\langle\Delta n\rangle_{C}=\bar{n}_{\mathrm{B}|C}-\bar{n}_{\mathrm{A}|C}$, we want arm B to have more photons. Due to the symmetry of the two modes, one can choose $\eta_{\mathrm{A}}>\eta_{\mathrm{B}}$ without loss of generality. It follows then both $\langle\Delta n\rangle_{(0,0)}$ and $\langle\Delta n\rangle_{(1,1)}$ are positive and $s(0,0)=s(1,1)=0$. With this optimal strategy, the expression of the mean photon number difference created by the demon with feed-forward is
\begin{equation}
\langle \Delta n\rangle=\frac{2\bar{n}(1-R)\widetilde{R}_{\mathrm{B}}[1+2\bar{n}\widetilde{R}_{\mathrm{A}}(2-\widetilde{R}_{\mathrm{A}})+\bar{n}^2(\widetilde{R}_{\mathrm{A}}^2(3-2\widetilde{R}_{\mathrm{A}})+\widetilde{R}_{\mathrm{A}}\widetilde{R}_{\mathrm{B}}(1-\widetilde{R}_{\mathrm{A}})^2)]}{(1+\bar{n}\widetilde{R}_{\mathrm{A}})^2(1+\bar{n}(\widetilde{R}_{\mathrm{A}}+\widetilde{R}_{\mathrm{B}}-\widetilde{R}_{\mathrm{A}}\widetilde{R}_{\mathrm{B}}))^2}\,.
\end{equation}
However, this expression is complicated so one cannot obtain an analytical expression for $\langle\Delta n\rangle_{\mathrm{max}}$ and the corresponding values for $R$ and the $\eta$'s. Instead, we numerically maximize this expression subject to the necessary constraints that the reflectivity and detector efficiencies are between 0 and 1, and Fig.~\ref{TMSSgraph} shows the result.

Finally, as a comparison with the other states, we note that when $\bar{n}=1$, we have $\langle\Delta n\rangle_{\mathrm{max}}\approx 0.272$, which is slightly larger than the value of $0.255$ obtained for uncorrelated thermal states. This is achieved with $R\approx 0.373$, $\eta_{\mathrm{B}}=1$ and $\eta_{\mathrm{A}}\approx 0.415$.
\end{widetext}

\end{document}